\documentclass[prb,twocolumn,showpacs,aps]{revtex4}
\usepackage {graphicx}
\pdfoutput=1
\begin{document}
\title{Cooperative ordering of superparamagnetic ZnO nanograins}

\author{B. Ghosh$^a$,M. Sardar$^b$, S. Banerjee$^a$\footnote{Email:sangam.banerjee@saha.ac.in}}

\address{$^a$ Surface Physics Division,  Saha Insitute of Nuclear Physics, 1/AF Bidhannagar, Kolkata 700 064, India,\\ $^b$ Material Science Division, Indira Gandhi Center for Atomic Research, Kalpakkam 603 102, India}

\begin{abstract}
In this paper we have tried to understand the precise nature of magnetism in ZnO nanoparticles. Cooling field dependence of magnetic hysteresis and coercive field was observed for high temperature annealed sample indicating cooperative magnetic correlation and ordering within the agglomerated nanograins. The increasing induced internal magnetic field along the direction of external field for the high temperature annealed sample has been fitted with the modified Weiss-Brillouin model indicating emergence of long range intergrain interaction among the superparamagnetic grains. We propose a simple idea that explains the reduction of magnetisation due to vortex state like flux-closure situation. 
\end{abstract}
\maketitle

Zinc oxide, is a multifunctional material that has various promising applications. Recently room temperature ferromagnetism (FM)
is widely reported in undoped ZnO nanoparticles\cite{our,gao,wang}, amine-capped nanoparticles\cite{amino},
 nanorods\cite{rod}, nanowires\cite{xing}, thin films\cite{film,xu}, carbon-doped ZnO\cite{carbon}, 
mechanically deformed ZnO powder\cite{mech}, and Zn-ZnO core-shell nanoparticles\cite{core}. 
FM was also observed in undoped wide-band gap semiconductors 
like, HfO$_2$, TiO$_2$, SnO$_2$, In$_2$O$_3$, Al$_2$O$_3$, CeO$_2$ etc\cite{others,sundar,spaldin}. 
There is a general agreement that magnetism in ZnO occurs from lattice defects like O vacancies\cite{our,gao,xing}, Zn
 vacancies\cite{wang,film}, Zn interstitial\cite{rod}. Theoretically Zn vacancy\cite{jena} was found to be responsible for moment formation. Earlier we\cite{our} had argued that O vacancy clusters (extended defect) are responsible 
for magnetic moment formation. These reports have created a lot of excitement and provoked us to do further experiment to understand the nature of magnetism in these pure semiconductors. Interestingly, FM is generally being claimed for these nanoparticles from the mere observation of  hysteresis in the magnetization (M) vs magnetic field (H) measured and this we believe might be misleading. Refined calculations have found that point defects such as Zn or O vacancy could not
 give rise to moment, but rather large extended defects (clusters etc.) can only be responsible\cite{extend} for moment 
formation in ZnO system. Similar arguments may be extended to explain the observation of magnetism in other 
(mentioned above) non-magnetic wide-band gap semiconductors too. Understanding the possible origin and investigating the 
precise nature of magnetism in ZnO now seems to be very important and challenging.
 Hysteresis in M vs H will always be observed below blocking temperature (which is greater than room temperature
in most cases mentioned above) if the nanoparticles have moments. Thus the claim, that most bulk diamagnetic non-magnetic semiconductors might be ferromagnetic in the nanoparticle form\cite{sundar} needs careful scrutiny. In this work, we shall present our investigation to understand the nature of magnetism with respect to particle size and multigrain (polycrystalline) formation of ZnO nanoparticle system. 

\begin{figure}
\includegraphics*[width=9cm]{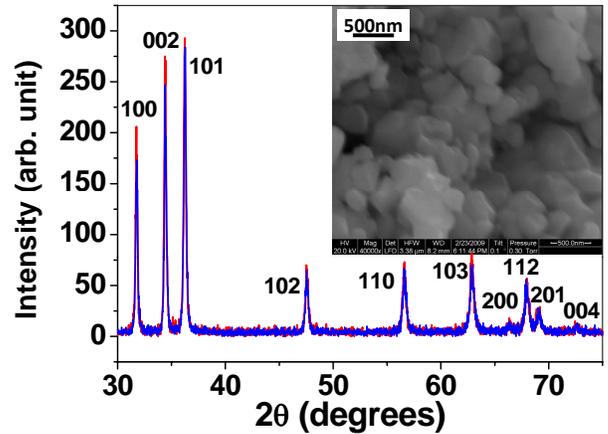}
\caption{{\bf (color online)} X-ray diffraction pattern of the 600$^0C$ and 900$^0C$ annealed samples (we see that both are overlapping). Inset:SEM micrograph showing the typical grain size 100 - 500 nm for the 900$^0C$ annealed sample}
\end{figure}

\begin{figure}
\includegraphics*[width=7cm]{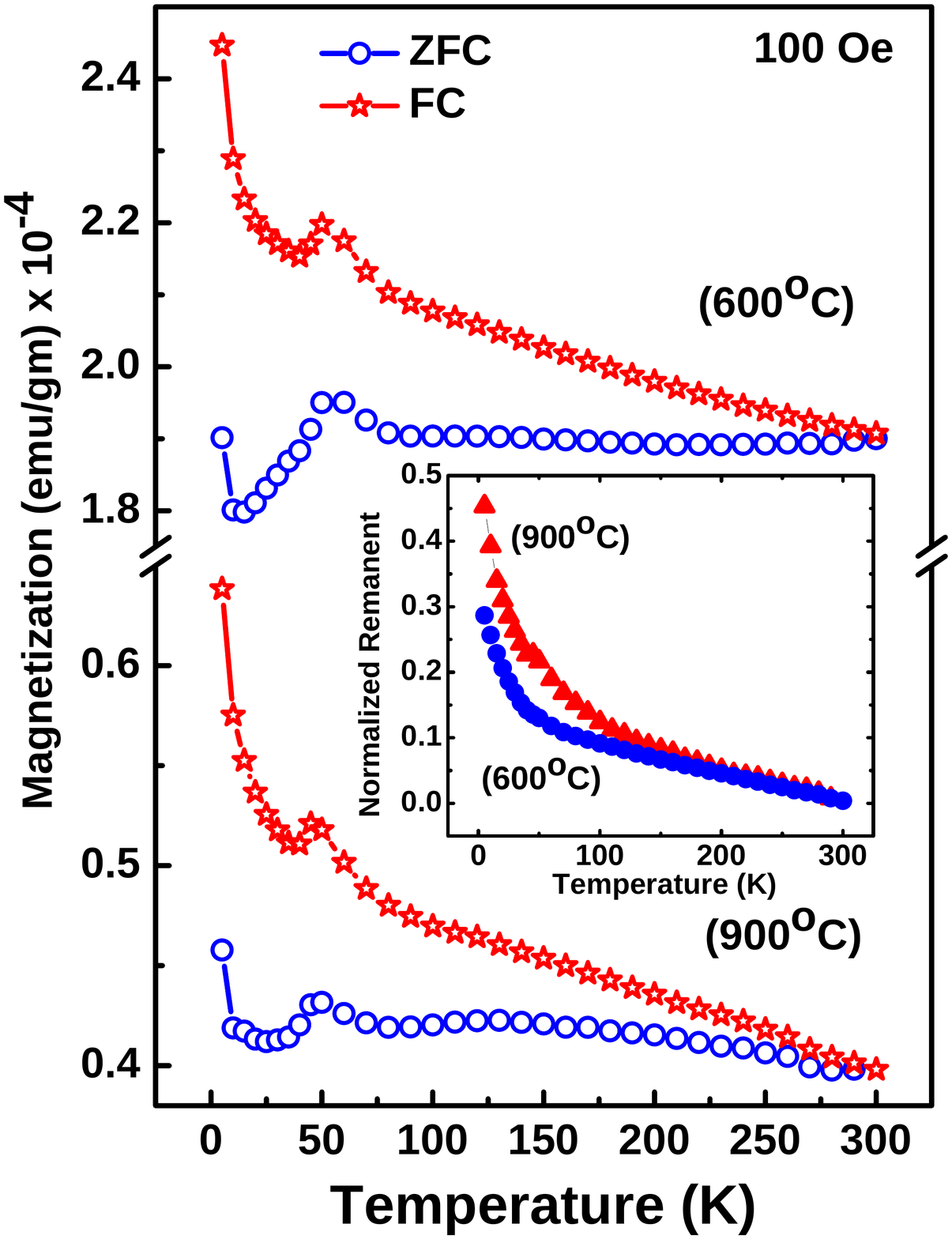}
\caption{{\bf (color online)} Zero field cooled (ZFC) and field cooled (FC) magnetisation M versus applied magnetic field H for (a) sample A (600$^0C$) and (b) sample B (900$^0C$), inset (c): Remanent magnetization}
\end{figure}

	ZnO nanoparticles were synthesized by solvo-thermal method \cite{synthesis} and then annealed at higher temperatures 600$^0$C (sample A) and 900$^0$C (sample B) to get high pure nanocrystalline ZnO samples. X-Ray Diffraction (XRD) data was in good agreement with standard diffraction data for zinc oxide and no impurity peak was observed for both the samples (see fig. 1). The intensity of the peaks is enhanced for sample B indicating increase of grain size. Average grain sizes calculated from XRD data using Debye-Scherer formula for sample A and B are 27 nm and 36 nm respectively. Scanning Electron Microscopy (SEM) exhibited agglomerated grains and the agglomerated particle size were $\sim$ 100 - 500 nm for both the samples. Hence, we could conclude that each agglomerated particles observed in SEM micrograph consists of many nano-grains having size observed using XRD. Energy Dispersive X-Ray Analysis (EDX) were also used to characterize the samples which shows the absence of any other chemical impurity. The magnetic property of both the samples were measured using MPMS-7 (Quantum Design).

\begin{figure}
\includegraphics*[width=7cm]{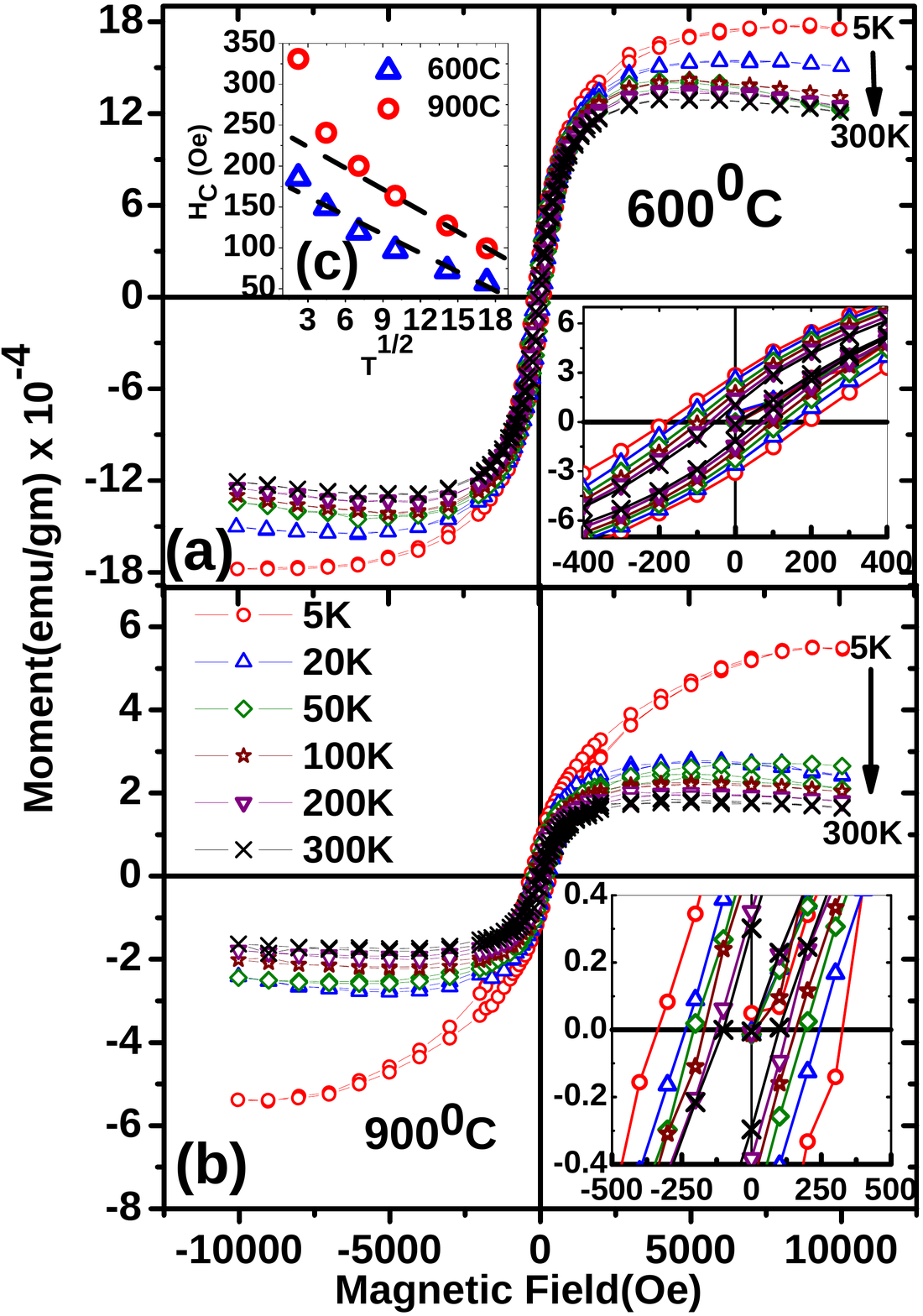}
\caption{{\bf(color online)} Magnetic hysteresis as a function of temperature for (a) sample A (600$^0C$) and (b) sample B (900$^0C$), inset: Coercive field vs. T$^{1/2}$}
\end{figure}

The main features of the magnetization study are as follows: (1) For both sample A and sample B the magnetization splits up above room temperatures as shown in fig.2. At $T< 10$ K there is a tendency of sharp increase in FC and ZFC magnetization, that looks like as due to a small paramagnetic contributions from isolated moments.
 The cusp in the ZFC and FC magnetization at 50K could be attributed to oxygen contamination on the surface \cite{MPMS} but 
recently some groups have attributed it to spin-glass behavior \cite{carbon,Lui}. 
Inset of fig.2 shows the remanent magnetisation ($M_{FC}-M_{ZFC}$ normalised by the magnetization at room temperature). There is a distinct break of slope at T= 50 K and 100 K for samples A and B respectively, which might be due to partial freezing of moments.
(2) Hysteresis is observed in both samples at all temperatures (see fig. 3(a,b)), which is usual if most of the moments are blocked.
(3) The saturation magnetisation at 7 Tesla of sample A is much higher ($2.2 \times 10^{-3}$ emu/gm ) than the sample B (
$5.0 \times 10^{-4}$ emu/gm) at room temperature (300K). (4) The coercive field of sample B is consistently higher than the coercive field of sample A at all temperatures (see fig.3(c)). At T=5K the coercive field is about 195 Oe and 335 Oe for samples A and B respectively. 
For noninteracting magnetic nanoparticles, generally $H_c \propto T^{1/2}$. The departure from $T^{1/2}$ law at low temperature is 
more pronounced for the sample B than the sample A as shown in inset fig.3(c). (5) In fig. 4 we see that the M-H loop of the sample A is 
independent of the external cooling field; whereas for sample B the cooling field dependence of M-H loop is spectacular. For sample B we see that with increase in cooling field the hystersis loop rotates counterclockwise and steady monotonic decrease of coercive field as shown in fig. 5(a). The magnetisation M$_{1T,FC}$ at 1T increases with cooling field (see fig.5(b)). Observation of very different magnetic behaviour of the two samples can bedue to the organization of microstructure and  the strength of the magnetic moments of the grain.

\begin{figure}
\includegraphics*[width=7cm]{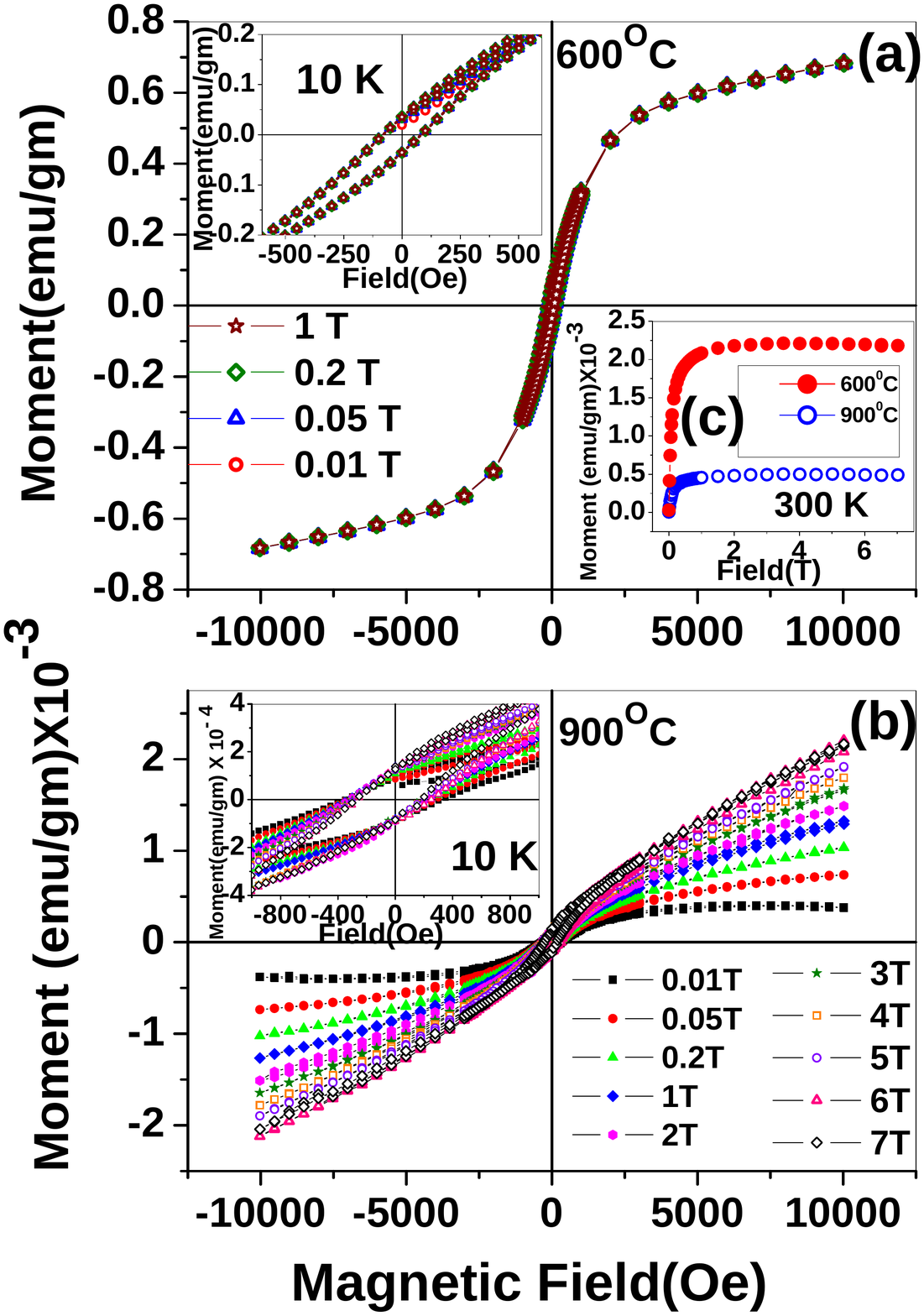}
\caption{{\bf(color online)} Magnetic hysteresis as a function of cooling field H$_{FC}$ for (a) sample A (600$^0C$) and (b) sample B (900$^0C$), inset (c) shows saturation magnetization at room temperature (300K).}
\end{figure}

\begin{figure}
\includegraphics*[width=7cm]{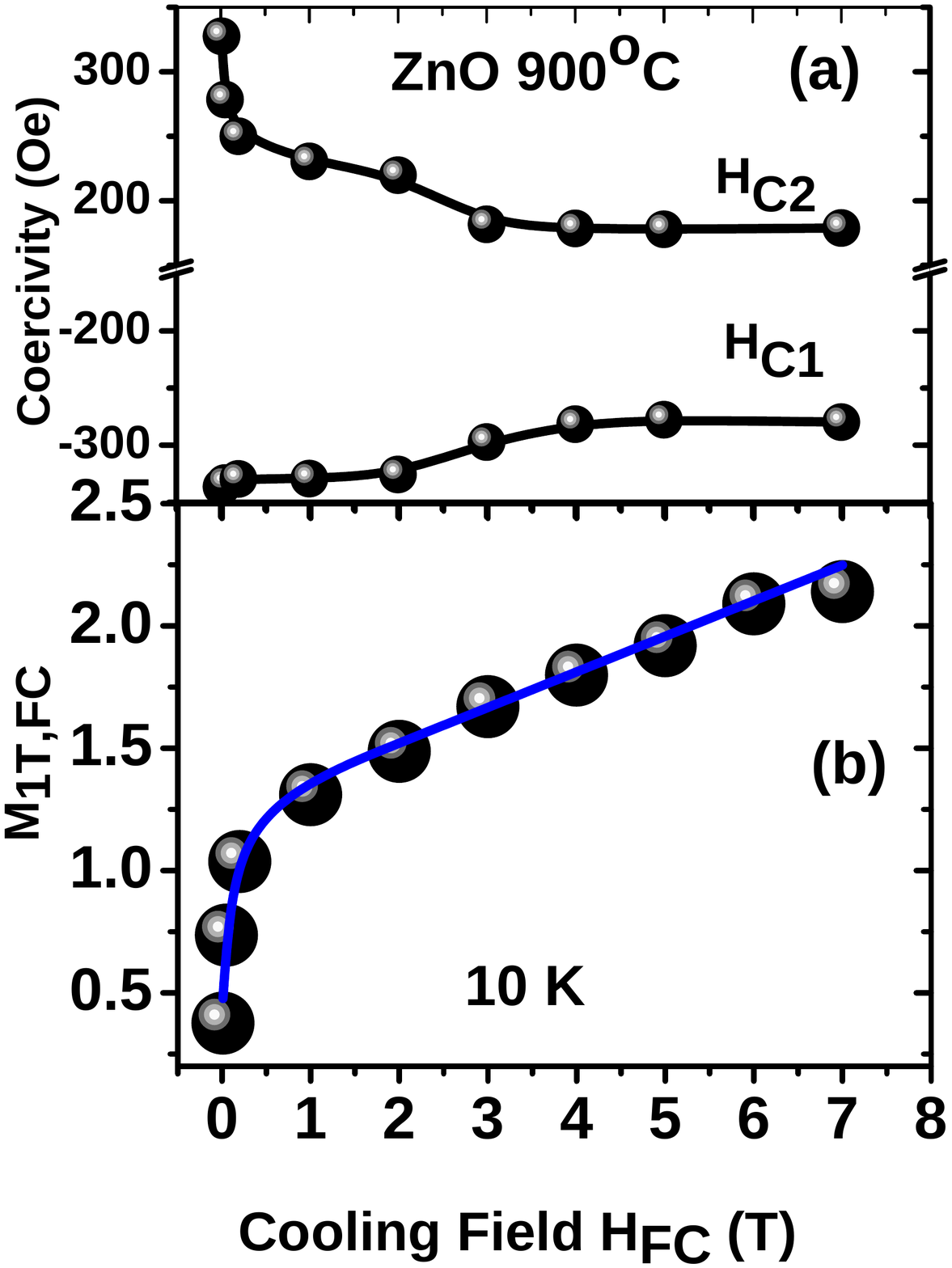}
\caption{{\bf(color online)} (a) Magnetic coercive field as a function of cooling field H$_{FC}$ and (b) Magnetization M$_{1T,FC}$ at H=1 tesla, as a function of cooling field H$_{FC}$ for sample B (900$^0C$), the solid line is fit to Eq. (1).}
\end{figure}

In fig.6 we show schematically the model proposed to explain our observation. The sample A, should be thought of as very weakly (dipolar type) interacting moments of the nanograins because each nanograins have very low magnetic moments as depicted in the model (small arrows - Fig.6)). These moments are blocked below room temperature, and that is why they show hysteresis. The lack of dependence of hysteresis loop shape on cooling field, is because the moments relax quickly once the cooling field is switched off. On the other hand upon higher annealing temperature (900$^0$C), the probability of clustering of vacancies increases for the sample B leading to increase in the moments of the grain (as indicated by big thicker arrows (fig.6)) along with the increase in grain size, similar to the result reported earlier \cite{our}. If the increase in the moment gives rise to emergence of intergrain magnetic interaction then the low energy state for such multigrain particles will be when the vacancy clusters on individual grains organise themselves to minimise the internal 
magnetostatic field energy. This will result into rearrangement of the local anisotropy directions (magnetic easy axis) to form a 
flux closure structure unlike the former sample A (schematically shown in fig.6(b)). That is why the saturation moment of sample B is less than the sample A. Frustrating dipolar interaction between the grains in the flux closure state will resist complete spin allignment of the multigrain particles, at low external field. The multigrain/multi-domain structure of 
the particles increases the coercivity of the sample B compared to sample A. External cooling field 
alligns these elementary moments in the grain along the external cooling field by rotating their 
easy axis because we observe increase of magnetization M$_{1T,FC}$ at H=1 tesla, as a function of cooling field.
The cooling field induced moment relax very slowly once the cooling field is switched off before measuring M-H loop at 10 K.
The progressive easy axis rotation along the cooling field direction leads to both anticlockwise rotation of
the hysteresis loops, and increase in M$_{1T,FC}$ with increase in cooling field.

\begin{figure}
\includegraphics*[width=8cm]{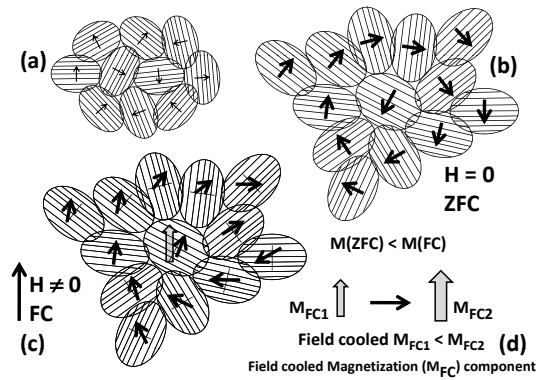}
\caption{Schematic representation showing each particle containing cluster of nano-grains and the strength of moments are shown in terms of thickness of the arrows (moments and size of grain in (a) is lower than in (b). In (c) the moments are rotated along the direction of external field, thin line with the arrows are the original moment direction. In (d) we show the strength of field-induced moments, for higher field-cooled samples the field-induced moments will be higher (shown as double arrow).}
\end{figure}

With this physical picture  we  can explain the cooling field dependence of M$_{1T,FC}$ by writing  the average moment of the grain ${\bar\mu}$ as ${\bar\mu}$ = ${\bar\mu_0}$ + ${\bar\alpha}H_{FC}$. The magnetization as a function of cooling field at 1 Tesla, M$_{1T,FC}$ can be expressed as

$M_{1T,FC} =M_{1T,ZFC} + N{\bar\mu}tanh\frac{\bar\mu H}{k_B T}$
\begin{equation}
=M_{1T,ZFC} + N({\bar\mu_0} + {\bar\alpha}H_{FC})tanh\frac{({\bar\mu_0} + {\bar\alpha}H_{FC})H}{k_B T} 
\end{equation}

\noindent where $M_{1T,ZFC}$ is the magnetization at H=1 Tesla, with zero field cooling, and is equal to $2.5\times 10^{-4}$ emu/gm.
The fit is excellent with parameters, $\mu_0= 2.36~\mu_B$, $N=4.11 \times 10^{16}$/gm, and $\alpha = 0.364~\mu_B$/Tesla. The average moments $\bar\mu$ varies from $2.38 \mu_B$ to $4.93 \mu_B$ as the cooling field is varied from zero to 7 Tesla. The average value of $\bar\mu$ is low due to flux-closure. The magnetic interaction between the superparamagnetic grains can be described by introducing Weiss molecular field \cite{Weiss} like term in paramagnetic Brillouin function and hence $\alpha H_{FC}$ term looks similar to Weiss's molecular field. $H_{FC}$ simply gives an additional effect of increasing the elementary moments along its direction and the field-induced moments (indicated as a double arrow in the schematic - fig.6) do not relax quickly as in sample A. This is because within each multigrain particle, the coupled dipolar interaction between the grain and the inhomogeneity of the easy axis of magnetization of all the defect clusters, resists fast relaxation. This is why with increasing cooling field the field cooled magnetization at 1T increases and the loop rotates anticlockwise. The coercivity monotonically decreases with increase in the external cooling field (fig.5(b)), because the cooling field-induced moments will follow the external applied magnetic field more easily indicating emergence of intergrain interaction among the superparamagnetic grains (long range) leading to vortex state like flux-closure situation. 

In conclusion, in this investigation we could point out that intergrain magnetic interaction among superparamagnetic nanograin in the agglomerated particles can only be observed if the moments of the grains are high. One observes flux-closure (vortex state) if the particles are having multigrain/multidomain structure. The magnetization M$_{1T,FC}$ follows the modified Weiss-Brillouin function indicating that the superparamagnetic grains have long range interaction. Reduction of coercivity as a function of external applied cooling field clearly also indicates  the emergence of intergrain magnetic interactions.

\end{document}